\documentclass[superscriptaddress,prl,twocolumn,aps,showpacs]{revtex4}
\usepackage{graphicx}
\usepackage{bm}
\begin{document}
\date{\today}

\title{Shell-Model Monte Carlo Simulations of Pairing in Few-Fermion Systems}
\author{N.T. Zinner}\email{zinner@phys.au.dk}
\affiliation{Department of Physics, Harvard University,
MA 02138, USA}
\affiliation{Lundbeck Foundation Theoretical Center for Quantum Systems Research, Department of Physics
and Astronomy, University of Aarhus,
DK-8000 {\AA}rhus C, Denmark}
\author{K. M\o lmer}
\affiliation{Lundbeck Foundation Theoretical Center for Quantum Systems Research, Department of Physics
and Astronomy, University of Aarhus,
DK-8000 {\AA}rhus C, Denmark}
\author{C. \"Ozen}
\affiliation{Department of Physics, Yale University, CT 06520, USA}
\affiliation{Gesellschaft f\"ur Schwerionenforschung GSI, D-64259 Darmstadt, Germany}
\author{K. Langanke}
\affiliation{Gesellschaft f\"ur Schwerionenforschung GSI, D-64259 Darmstadt, Germany}
\affiliation{Technische Universit\"at Darmstadt, D-64289 Darmstadt, Germany}
\author{D.J. Dean}
\affiliation{Physics Division, Oak Ridge National Laboratories, Oak Ridge, TN 37831, USA}

\begin{abstract}
We study a trapped system of fermions with an attractive zero-range two-body
interaction using the Shell-Model Monte
Carlo method. The method provides {\em ab initio} results in the low
$N$ limit where mean-field theory is not applicable. The
energy and pairing properties are presented as functions of
interaction strength, particle number, and temperature. In the
interesting region where typical matrix elements of the two-body
interaction are comparable to the level spacing of the trap we find
large odd-even effects and signatures of shell structure. As a
function of temperature, we observe the disappearance of these
effects as in a phase transition.
\end{abstract}

\maketitle

The physics of ultracold atomic gases has been intensively pursued
experimentally and theoretically in the last decade. Recently there
has been great interest in strongly-interacting Fermi
gases where Feshbach resonances allow the tuning of the two-body interaction.
Studies of the transition
from a dilute gas of fermionic atoms to a Bose condensate of
molecules have therefore been possible in the laboratory
\cite{ohara2002,dieckmann2002,chin2004,zwierlein2005,kinast2005,stewart2006}.
While studies of degenerate Fermi gases have mostly dealt with large
atom numbers and wide traps, efforts have begun to trap only few
(1-100) atoms in tighter traps \cite{jochim2008}. Also, with the
implementation of three-dimensional optical lattices, a
low-tunneling regime can be reached with essentially
isolated harmonic oscillators containing only a few fermions at
each site \cite{stoferle2006}. This means that one can now
explore few-body fermionic effects in trapped systems with
scattering lengths that are comparable to the inter-particle distance
and the trap width.

In this {\it letter} we report on a theoretical study of harmonically
trapped fermions using the Shell-Model Monte Carlo (SMMC) approach.
This method has been extensively used in nuclear physics to
determine nuclear properties at finite temperature in larger model
spaces than can be handled in normal nuclear shell-model
diagonalization \cite{johnson1992,koonin1997}. In the SMMC the
many-body problem is described by a canonical ensemble at
temperature $T=\beta^{-1}$ and the Hubbard-Stratonovich
transformation is used to linearize the imaginary-time many-body
propagator $e^{-\beta H}$. Observables are then expressed as path
integrals of one-body propagators in fluctuating auxiliary fields.
The method is in principle exact and subject only to statistical
uncertainties.
For equal mixtures of two hyperfine
states at low density, the interaction can be modeled
with an $s$-wave zero-range potential. It thus belongs to the
class of two-body interactions which is free of sign problems in the
SMMC \cite{koonin1997}. We present here the first application of this many-body
method to ultracold gas physics.

We study the usual zero-range pairing Hamiltonian
\begin{equation}
H=\sum_i \frac{p_{i}^{2}}{2m}+\sum_i \frac{1}{2}m\omega^2 r_{i}^{2}+\sum_{[ij]}V_0
\delta(\vec{r}_i-\vec{r}_j),
\end{equation}
where we sum over all particles $i$ and $[ij]$ denotes a sum over
fermion pairs with opposite internal states. The trap frequency is
$\omega$, and $V_0$ denotes the interaction strength.
The SMMC method was originally set up to handle nucleons, and pairing
problems have been extensively considered \cite{dean2003}. The spin symmetric
ultracold Fermi gas can now be easily mapped onto a single spin $1/2$ nucleon species \cite{heiselberg03}, as
the two-body interaction in the Hamiltonian corresponds to standard
nuclear pairing.

Simple dimensional arguments reveal that the matrix elements scale as
$1/b^3$, where $b=\hbar/m\omega$ is the oscillator length. It is therefore natural
to redefine the interaction strength in terms of $V_0=-g \hbar\omega
b^3$, where $g$ is a dimensionless strength measure. To relate
$V_0$ to the $s$-wave scattering length, one needs to regularize the zero-range potential
and consider the finite
model space cut-off, as pointed out in \cite{weinberg1991}.
Here we will adopt the treatment in \cite{esbensen1997} and renormalize through 
continuum low-energy scattering.
The energy cut-off used can be written $E_c=\alpha^2\hbar\omega$
with $\alpha^2=N_{max}+3/2$. This defines a relation between $V_0$, $a$, and $E_c$, which 
can be written as $4\pi a/b=g/(\alpha g/2\pi^2-1)$. 
This gives the effective interaction
strength needed for a model space with $N_{max}+1$ major shell.
The relevant parameter regime is
expected to be where the natural energy scale, given by the level spacing $\hbar\omega$, 
is comparable to typical
two-body matrix elements between the trap states at $T=0$ (as a quantitative measure we use the
$(1s)^2-(1s)^2$ element). This turns out to be around $g\sim 10$ in
our setup, which corresponds to $a=10.66 b$. The different
regions of interest in terms of interaction strength and
$\hbar\omega$ in the context of atomic gas
physics are discussed in \cite{heiselberg2002,bruun2002}.
We note, however, that our procedure to relate $g$ and $a$
is only approximate
and  modifications to the simple relation we use can occur, e.g.  
in tight traps \cite{bolda2002}.

Previous works have considered few-fermion systems using advanced many-body methods. The Green's function
Monte Carlo
methods were applied to homogeneous \cite{carlson2003}, as well as trapped systems \cite{bertsch2007}.
No-core \cite{stetcu2007}, and traditional shell-models \cite{alhassid2007}, using effective interactions
have also recently been applied to these systems, particularly for very low particle numbers where exact
results are available \cite{werner2006a,werner2006b}. Finite-temperature, non-perturbative lattice methods
have
also been applied to homogeneous \cite{bulgac2007} and trapped fermions \cite{akkineni2007}. 
These works focus on 
the unitary $|a|\rightarrow \infty$ limit and the crossover regime around it.
The present SMMC approach is not variational, but an exact many-body method
subject only to statistical
uncertainties. 
Our choices of $N_{max}$ and $g$ discussed above imply that the SMMC results given here
are on the BEC side of the crossover regime.

\begin{table}
\caption{Energies (in units of $\hbar\omega$) calculated with the SMMC method for a trapped fermion gas
with interaction parameter
$g=10$ calculated at temperatures (in units of $\hbar\omega$)
$T=1/6$ and
$T=1/5$ for different particle numbers $N$. The statistical uncertainty is given in parenthesis.
HOSD denotes the non-interacting energies at $T=0$.}
\begin{tabular}{|c|ccc||c|ccc|}
\hline
N  & HOSD & $T=1/6$ & $T=1/5$   &   N & HOSD & $T=1/6$ & $T=1/5$  \\
\hline
2  & 3  & 2.60(1) & 2.62(1)  & 12 & 32 &29.8(2)&29.9(1)    \\
3  & 5.5 & 5.12(1)& 5.14(1)  & 13 & 35.5 & 32.8(4)&33.3(2)    \\
4  & 8    & 7.31(1)& 7.37(1)    & 14 & 39 &36.4(3)&36.6(2) \\
5  & 10.5 &9.73(2)&9.76(1)    & 15 & 42.5 &39.8(2)&39.9(2)  \\
6  & 13& 11.9(1)&11.9(1)      & 16 & 46 &43.0(2)&43.1(2)   \\
7  & 15.5 &14.2(2)&14.3(1)   & 17 & 49.5 &46.3(2)&46.5(2)  \\
8  & 18 &16.3(1) &16.4(1)    & 18 & 53 &49.5(2)&49.7(2)  \\
9  & 21.5 &19.8(1)&19.9(1)   & 19 & 56.5 &52.9(2)&53.1(2)   \\
10 & 25 &23.1(2)&23.2(1)     & 20 & 60 & 56.1(2)&56.4(1)\\
11 & 28.5 & 26.5(2)&26.6(1)    &    &  &  \\
\hline
\end{tabular}
\label{g10tab}
\end{table}

In Tab. \ref{g10tab} we present the SMMC energies for two different
temperatures small compared to $\hbar\omega$, along with the
non-interacting energies.
There is agreement between the two SMMC calculations within
the uncertainties for almost all particle numbers. Notice that the
absolute energy uncertainty decreases with increasing $T$, as
expected \cite{koonin1997}.
The $N=2$ problem can be solved exactly for all $a$ \cite{busch1998}, and with $a=10.66b$ we find 2.4
for the $T=0$ ground-state.
Our value of 2.6 at $T=1/6 \hbar\omega$ thus indicates that we are close to the $T=0$ limit.
In general, we find that our
energies are typically above those calculated at unitarity in \cite{werner2006a,bertsch2007}
and in the crossover regime in \cite{blume2008}. This is expected due to finite temperatures and
also because of Hilbert space differences (see discussion below).

\begin{figure}
\begin{center}
\includegraphics[clip=true,scale=0.29]{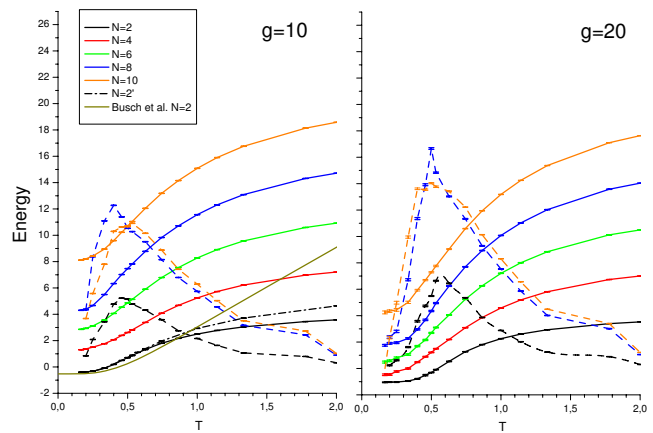}
\caption{(color online) The temperature dependence of the SMMC energies 
(both in units of $\hbar\omega$) for selected even $N$ systems 
and for strength parameters  $g=10$ (left) and $20$ (right). The zero-point
energies are subtracted.
Statistical uncertainties are too small to show.
The dashed curves give the specific heat
for $N=2,8,$ and $10$.
Also shown are the exact $N=2$ results 
(Busch {\it et. al.}, \cite{busch1998}) with $a=10.66b$ for the full space.
The dash-dotted curve $N=2'$ gives the SMMC results for 5 major shells 
with $g=8.98$.}
\label{fig1}
\end{center}
\end{figure}

Fig. \ref{fig1} shows the SMMC energies as a function of temperature
for selected even particle
numbers and for the strength parameters $g=10$ and $20$.
We can clearly see the
convergence of our results at small $T$,
approaching the T=0 ground state value.
Quantitatively we find that the energies increase by about 1\% (2\%)
going from $T=\hbar\omega/6$ to $T=\hbar\omega/5$ ($T=\hbar\omega/4$).
At high $T$, the curves flatten out as they approach the equipartition limit of equal
occupancies in the finite model space.

We have used moderately small model spaces of 4 major shells for most
calculations, which might be problematic when using zero-range 
interactions. However, to check our convergence we have repeated the 
calculation for N=2 in an enlarged model space of 5 major shells with the 
interaction strength readjusted to the same scattering length parameter.
As can be seen in Fig. 1 the two calculations, performed for $g=10$ for 
$N_{max}=3$ and $g=8.98$ for $N_{max}=4$, agree within the statistical 
uncertainty up to a temperature
$T \approx 0.7$. We also compare our SMMC results with the exact solution 
of the N=2 system as given in Ref. [25]. The comparison, although quite
reasonable at low temperatures, can only be indicative as the
pseudopotential used in [25] is similar, but not identical to the present
delta interaction and allows for a 2-body $1/r$ bound state which our
calculation does not include. At large T, the energy expectation value
approaches $E(T) = 3 N k_B T$ in the full model space. This behavior is,
however, not obtained in our restricted spaces where $E(T)$ converges to
the thermal equilibrium values which are 4.5 and 6 for $N_{max}=3$
and 4, respectively.

The $g=10$ results show
a sharp increase in the energy around $T \sim 0.3$.
Correspondingly the specific heat $C=dE/dT$ exhibits
a distinct peak around this temperature. Such a peak in the specific heat
is usually associated with phase
transitions in finite systems
and indicates that the system changes its
character. For $g=20$ we see the peak at higher $T$ as expected.
As we will show below,
it can be interpreted as the analogue of the superfluid-to-normal
phase transition in finite
nuclear
systems \cite{dean1995,langanke2005} and is
associated with the breaking of pairs. Notice in particular that the peak
height is larger for $N=8$ than $N=10$. This is a result of the shell closure at $N=8$
which causes the energy to be released in a smaller interval as $T$ increases.
The sizable width of the peak is due to the finite size of the system. 
The model space discussion above demonstrates that our peaks 
are genuine and not a finite size Schottky effect \cite{koonin1997}.

To relate the peak in the specific heat to changes in the
pairing correlations, we have calculated the expectation value of a
number-conserving BCS-like pair matrix
\begin{equation}
M_{\alpha,\alpha'}=\langle \Delta^\dagger(j_a,j_b)\Delta(j_c,j_d)\rangle,
\label{pairmat}
\end{equation}
with the $J=0$ pair operator
\begin{equation}
\Delta^\dagger=\frac{1}{ \sqrt{1+\delta_{ab}} }
\left[ a_{j_a}^{\dagger} \times a_{j_b}^{\dagger} \right]^{JM=00}
\end{equation}
where $a_{j_a}^{\dagger}$ creates a particle in orbit $j_a$ (which is the combination of orbital and spin
angular momentum of the fermions). This operator is thus a measure of the pairs with $J=0$.
An indication of the pairing correlations can be
obtained from the sum over all matrix elements, called
pairing strength in the following \cite{ozen2007}.
Since we are only interested in
genuine pair correlations we subtract the 'mean-field' values
obtained at the same $T$ but with $g=0$.

\begin{figure}
\begin{center}
\includegraphics[scale=0.29,clip=true]{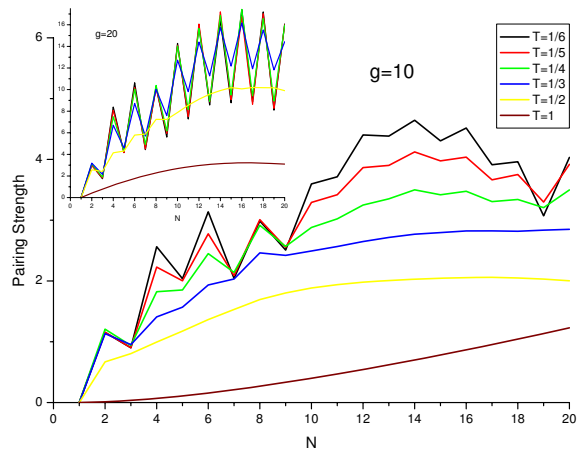}
\caption{(color online) The pairing strength as a function of particle number, $N$, for various
temperatures, $T$ (in units of $\hbar\omega$) for $g=10$. The upper left inset shows the
results for $g=20$. The uncertainties are very small and not shown.}
\label{fig2}
\end{center}
\end{figure}

In Fig. \ref{fig2} we show the pairing strength as a function
of particle number for different temperatures.
The most striking feature is naturally the odd-even staggering. The
relative reduction of pairing strength for odd-particle numbers is related
to the blocking of scattering of pairs into the orbital occupied by the
unpaired particle. The non-interacting systems have closed-shell configurations
for $N=2,8,20$. With interaction switched on, these configurations manifest
themselves by a relative reduction of the pairing strength (overlaid
by a general increase due to a growing number of pairs) and a larger
resistance against temperature increase. The strong dips observed for
particle numbers $N=7,9$ and 19 are also related to the shell closures.
Relatedly the pairing strength is largest for mid-shell systems. From the inset
we see that the staggering is larger for $g=20$ and persists to larger
temperatures as expected.

As function of temperature, the pairing
strength for $g=10$ decreases and the odd-even effects vanish around
$T\sim(1/3-1/2)\hbar\omega$ except for the lowest $N$.
This is consistent with the value at which the energies show sharp increases in Fig. \ref{fig1} with
resulting peaks in the specific heats. The same is seen for the $g=20$ results.
At higher $T$ we see only a monotonous increase with $N$, indicating
equipartition in the model space with loss of all shell structure.

\begin{figure}
\begin{center}
\includegraphics[scale=0.29,clip=true]{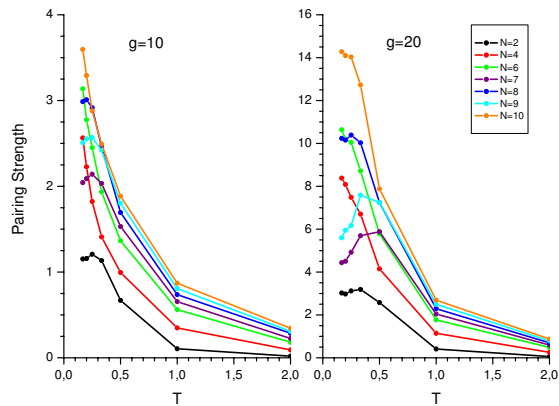}
\caption{(color online) The pairing strength 
as a function of temperature, $T$ (in units of $\hbar\omega$), for various
particle numbers, $N$ and for $g=10$ (left) and $g=20$ (right). 
Uncertainties are small and not shown. Note the different  scales
in the two panels.}
\label{fig3}
\end{center}
\end{figure}

To investigate further the transition between a paired state
and a normal state, we show the pairing
strength for $g=10$ and $g=20$ as a function of $T$ for selected particle numbers in Fig. \ref{fig3}.
We note that the
pairing correlations decrease rapidly for all particle numbers in the
temperature regime which corresponds to the peak structure in the
specific heat (with the strong decrease happening at larger $T$ for $g=20$ as expected).
We
have calculated the pairing strength up to $T=4$, at which point it has practically vanished for all
$N$. We believe this indicates that the
transition temperature for our low particle numbers is some numerical
factor less than unity (depending on $g$ and $N$) times the trap spacing $\hbar \omega$.
Notice how the closed shell numbers $N=2$ and 8 have different
low-$T$ behavior from the other systems
(also the case for $N=20$ which is not shown).
Here the pairing remains relatively constant to larger temperatures
due to the shell closure. Shown are also $N=7$ and $N=9$
which exhibit the large dips in the pairing strength
in Fig. \ref{fig2}. They are generally below the neighboring even-$N$ systems at low $T$, yet again
confirming that the unpaired particle has a significant
blocking effect on the pairing strength.
Furthermore they have the same structure as the neighboring closed
shell $N=8$, but at lower
magnitudes. It is also interesting to observe that, for $N=7$ and $9$, the
pairing strength is largest at finite $T$ reflecting the competition
between blocking by the unpaired particle and thermal excitations
which moves the unpaired particle across the shell closure reducing
the blocking effect. A similar effect has been found in SMMC
studies for nuclei with odd nucleon numbers \cite{Engel98}.
At higher $T$ these effects are smoothed out as pairing vanishes
and the pairing strength orders with increasing number of particles.
Comparing the $g=10$ and $g=20$ results we see that the above effects are
more pronounced for the stronger pairing strength and persists to higher temperature.
This is consistent with the discussions above. Similar evidence for a transition at finite $T$ in 
the homogeneous system in both energy and pair correlation was found in \cite{akkineni2007}.

We have also considered the regime of $g<10$ and found that at $g=1$
(deep BCS regime) there are basically none of the interesting effects left. We
therefore predict that for small systems with simple two-body
pairing, there is a regime at $a>b$ where pairing features are
pronounced. For $g>20$ the calculation becomes unstable, 
so we cannot access this deep BEC regime. 

In conclusion, the SMMC offers a good quantitative description of
the behavior of equal mixtures of fermions in an interesting
interaction regime. We studied the case of isotropic traps, and we
extracted a number of particularly relevant properties. The
excellent convergence properties of the method holds promise for
application to specific situations where, e.g., deformation
properties and formation of higher angular momentum pairs may become
relevant. Deformation was studied in the nuclear case using the SMMC
to predict shape transitions occurring as a function of temperature in 
the competition between pairing and quadrupole interactions \cite{koonin1997}.
This is relevant also for ultracold gases with the recent realization 
of condensates with intrinsic long-range interactions between the atoms
\cite{lahaye2007}. Recently the SMMC was also used to study the parity and 
spin properties of the density of states in nuclear systems \cite{alhassid2007b,kalmykov2007}.
Transforming this to the atomic system could help us understand the
low-energy excitation spectrum and the response of the gas to external perturbations.

The authors wish to thank A.S. Jensen, M. Th\o gersen, and A.
Juodagalvis for many fruitful discussions.

\end{document}